\documentstyle[twoside,fleqn,espcrc3,epsf]{article}

\def\bu{B^+}
\def\bd{B^0_d} 
\def\bs{B^0_s}
\def\bmix{B^0 \mbox{--} \overline{B^0}}
\def\bdmix{B_d^0 \mbox{--} \overline{B_d^0}}
\def\bsmix{B_s^0 \mbox{--} \overline{B_s^0}}
\def\Zbb{Z^0 \rightarrow b{\overline b}}

\def\str{\penalty-10000\hfilneg\ } 

\newcommand{\AmS}{{\protect\the\textfont2
  A\kern-.1667em\lower.5ex\hbox{M}\kern-.125emS}}

\begin{document}

\pagestyle{empty}

\title{
\begin{flushright}
{\normalsize
SLAC--PUB--7567\\
June 1997\\}
\end{flushright}
      {\large\bf $B$ Physics at SLD}
      }

\author{
        S. Willocq\address{Stanford Linear Accelerator Center \\
        Stanford University, Stanford, CA 94309, USA \\ ~~\\
        Representing the SLD Collaboration$^{\ast}$ \\
        Stanford Linear Accelerator Center \\
        Stanford University, Stanford, CA 94309, USA \\
       }%
       }

\begin{abstract}

~~~
\vspace{10mm}
~~~
\hfill\break

{\normalsize
We review recent $B$ physics results obtained
in polarized $e^+ e^-$ interactions at the SLC
by the SLD experiment.
The excellent 3-D vertexing capabilities of SLD are exploited
to extract precise $\bu$ and $\bd$ lifetimes, as well as
measurements of the time evolution of $B^0_d - \overline{B^0_d}$ mixing.
}
\begin{center}

\vspace{65mm}
{\normalsize\sl
     Presented at the 5$^{th}$ Topical Seminar on The Irresistible Rise of
     the Standard Model,
     21-25 April 1997, San Miniato al Todesco, Italy.}

\vspace{20mm}
{\footnotesize Work supported in part by DOE Contract
                DE-AC03-76SF00515(SLAC).}
\end{center}

\end{abstract}

\maketitle

\pagestyle{plain}
~~~~
\vspace{5cm}
~~~~
\vfill\eject
~~~~
\vspace{5cm}
~~~~
\pagebreak

\section{INTRODUCTION}

The results presented here are based on samples of
approximately 50,000 and 100,000
$e^+ e^- \rightarrow Z^0 \rightarrow {\mbox{hadrons}}$ events
with longitudinally polarized electrons
collected by the SLD experiment during the 1993 and
1994--95 data taking periods, respectively.
During these periods, the average measured polarization was
$(63.0 \pm 1.1)\%$ and $(77.2 \pm 0.5)\%$.
A description of the detector can be found in Ref.~\cite{Rb}.
$\bu$ and $\bd$ lifetime results are presented in Sec.~\ref{Sec_life}
and $\bdmix$ mixing measurements are discussed in Sec.~\ref{Sec_bmix}.

\section{$\bu$ AND $\bd$ LIFETIMES}
\label{Sec_life}

  Measurements of the $B$ hadron lifetimes
are important to test our understanding of $B$ hadron decay dynamics.
In the naive spectator model, one expects
$\tau(\bu) = \tau(B^0_s) = \tau(\bd) = \tau(\Lambda_b)$.
However, a strong hierarchy is observed in the case of charm hadrons:
$\tau(D^+)\simeq 2.3~\tau(D_s)\simeq 2.5~\tau(D^0)\simeq 5~\tau(\Lambda_c^+)$.
This hierarchy is predicted to scale with $1/m_Q^2$ and is thus
expected to yield much smaller lifetime differences for $B$ hadrons.
A calculation~\cite{Bigi}, based on an
expansion in terms of $1/m_Q$,
predicts the lifetimes for different $B$ hadrons to be less than 10\%.

  Several techniques have been used to measure the $\bu$ and $\bd$
lifetimes. The cleanest method reconstructs samples of $\bu$ and
$\bd$ decays exclusively but suffers from small branching fractions
($\sim 10^{-4} - 10^{-3}$).
Most measurements have relied on samples of semileptonic $B$ decays where
the $D^{(\ast)}$ meson is exclusively reconstructed and intersected
with an identified lepton to determine the $B$ decay point.
  The two techniques used by SLD take advantage of the excellent 3-D
vertexing capabilities of the CCD Vertex Detector to reconstruct the decays
(semi-) inclusively. The goal is to reconstruct and identify all the
tracks originating from the $B$ decay chain, and thus
separate charged and neutral $B$ mesons using
the total charge $Q$ of tracks associated with the decay.

  The first analysis~\cite{blife} uses an inclusive topological vertexing
technique~\cite{ZVTOP} summarized below.
A search is made to find regions
in 3-D space with high track density (other than the primary vertex).
Such a region, or ``seed'' vertex, is found in $\sim 50\%$
of $b$ hemispheres, but only in $\sim 15\%$ of $c$~hemispheres
and in less than 1\% of $uds$ hemispheres.
The $b$~hemisphere vertex finding efficiency increases with
the decay length $D$ to attain a constant level of 80\% for $D > 3$~mm.
Due to the typical $B \rightarrow D$ cascade structure of the decays,
not all tracks originate from a single space point, and thus, may not
be attached to the seed vertex if the $D$ meson travelled sufficiently
far from the $B$ decay point.
Therefore, isolated tracks with $T < 1$~mm and $L/D > 0.3$
are attached to the seed vertex to form the final secondary vertex.
The quantity $T$
represents the minimum distance between a given
track and the seed vertex axis, and $L$ is the distance along the
vertex axis between the interaction point and the point of
\str\noindent
closest approach between the track and the vertex axis.

In the hadronic $Z^0$ event sample, we select 9719 $B$ decay candidates
by requiring the decay length $D > 1$ mm and the invariant mass computed
using all tracks associated with the
secondary vertex $M_{raw} > 2$ GeV/c$^2$.
The minimum vertex mass requirement serves not only as a means to select
a 97\% pure sample of $b\bar{b}$ events but also as a means to enhance
the charge reconstruction purity.
The sample is divided into 3665 neutral and 6033 charged
decays with $Q = 0$ and $Q = \pm 1, 2, 3$, respectively.
Monte Carlo (MC) studies show that
the charged sample consists of 52.8\% $\bu$,
32.1\% $\bd$, 8.6\% $B_s^0$, and 4.3\% $B$ baryons,
whereas the neutral sample consists of 
25.3\% $\bu$, 52.9\% $\bd$, 13.9\% $B_s^0$,
and 6.2\% $B$ baryons\footnote{Reference
to a specific state
(e.g., $B^+$) implicitly includes its charge conjugate state (i.e., $B^-$).}.
The sensitivity of this analysis to the individual $\bu$ and
$\bd$ lifetimes can be assessed from the
1.6 (2.1) ratio of $\bu\:(\bd)$ decays over $\bd\:(\bu)$ decays
in the charged (neutral) sample.

\begin{figure}[htb]
  \vspace{0mm}
  \centering
  \epsfxsize7.4cm
  \leavevmode
  \epsfbox{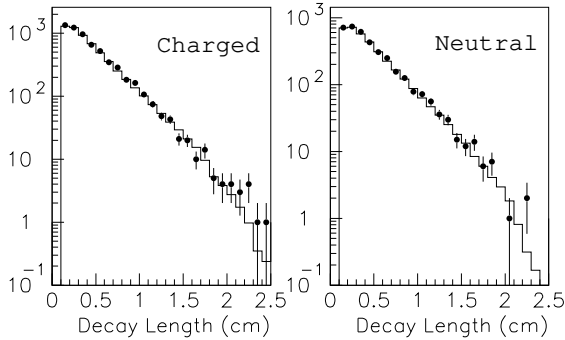}
  \vspace{-6mm}
  \caption{Decay length distributions for data (points) and best fit
           MC (histogram) in the topological analysis.}
  \label{Fig_dkltop}
\end{figure}

  The $\bu$ and $\bd$ lifetimes are extracted with a simultaneous
binned maximum likelihood fit to the decay length distributions of
the charged and neutral samples (Fig.~\ref{Fig_dkltop}).
These distributions are compared with MC distributions obtained for
a range of values of the $\bu$ and $\bd$ lifetimes.
The maximum likelihood fit yields lifetimes of
 $\tau_{\bu} = 1.67\pm 0.07$(stat) $\pm~0.06$(syst) ps,
 $\tau_{\bd} = 1.66\pm 0.08$(stat) $\pm~0.08$(syst) ps,
with a ratio of
 $\tau_{\bu}/\tau_{\bd} = 1.01^{+0.09}_{-0.08}$(stat) $\pm~0.05$(syst).
The main contributions to the systematic error come from uncertainties
in the detector modeling, $\bs$ lifetime, fit systematics, and MC statistics.

  The second lifetime analysis~\cite{blife} is restricted
to semileptonic decays. This reduces the overall efficiency compared to the
topological method but results in an improved charge reconstruction purity.
In this method, a $D$ decay vertex is reconstructed topologically
and the $B$ decay vertex is formed by intersecting the $D$ meson
trajectory with that of an identified lepton.
An attempt is then made to attach a slow-pion candidate to the $B$ vertex
to reconstruct the track topology of $B$ decays into $D^{\ast +}$ mesons.

  The analysis selects identified electrons and muons with
momentum transverse to the nearest jet axis $> 0.4$ GeV/c and results in
a sample of 634 charged and 584 neutral decays.
MC studies show that
the remaining charged (neutral) sample is 97.4\% (98.9\%) pure in
$B$ hadrons. The simulated flavor contents are 66.6\% $\bu$,
22.9\% $\bd$, 5.5\% $\bs$, and 2.4\% $B$ baryons for the charged
sample, and
19.5\% $\bu$, 60.2\% $\bd$, 14.8\% $\bs$, and 4.4\% $B$ baryons
for the neutral sample.
The sensitivity of this analysis to the individual $\bu$ and
$\bd$ lifetimes can be assessed from the
3:1 ratio of $\bu\:(\bd)$ decays over $\bd\:(\bu)$ decays
in the charged (neutral) sample.

\begin{figure}[t]
  \vspace{-7 mm}
  \centering
  \epsfxsize8cm
  \leavevmode
  \epsfbox{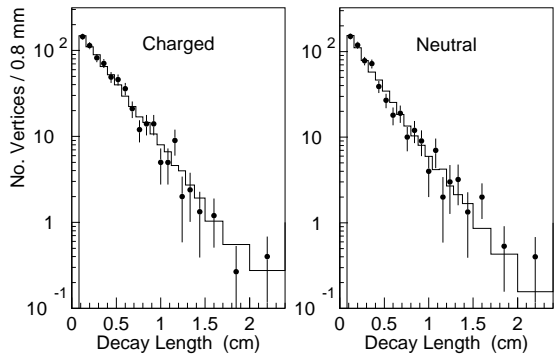}
  \vspace{-8mm}
  \caption{Decay length distributions for data (points) and best fit
           MC (histogram) in the semileptonic analysis.}
  \label{Fig_dklslep}
\end{figure}

  As for the topological analysis, the $\bu$ and $\bd$ lifetimes are
extracted from the decay length distributions of the charged and neutral
samples (Fig.~\ref{Fig_dklslep}).
The fit yields:\vfill\pagebreak\noindent
 $\tau_{\bu} = 1.61^{+0.13}_{-0.12}$(stat) $\pm~0.07$(syst) ps,
 $\tau_{\bd} = 1.56^{+0.14}_{-0.13}$(stat) $\pm~0.10$(syst) ps,
with a ratio of
 $\tau_{\bu}/\tau_{\bd} = 1.03^{+0.16}_{-0.14}$(stat) $\pm~0.09$(syst).
The dominant sources of systematic error are the same as for the
topological analysis.

The two analyses described above yield lifetime measurements in
agreement with those from other experiments
and with the expectation that the $\bu$ and $\bd$ lifetimes
are nearly equal.

\section{$\bmix$ MIXING}
\label{Sec_bmix}

  Transitions between flavor states $B^0 \rightarrow \overline{B^0}$
take place via second order weak interactions ``box diagrams.''
As in the case of the $K^0 - \overline{K^0}$ system, the weak
eigenstates are linear combinations of the flavor eigenstates.
Due to the difference in mass between the weak eigenstates, they
propagate differently in time, which gives rise to time-dependent
oscillations between $B^0$ and $\overline{B^0}$ flavor eigenstates.
The oscillation frequency $\Delta m_d$ for $\bdmix$ mixing
depends on the Cabibbo-Kobayashi-Maskawa (CKM) matrix element
$\left| V_{td} \right|$ for which little is known experimentally.
Theoretical uncertainties~\cite{Gibbons} are significantly reduced
for the ratio between $\Delta m_d$ and $\Delta m_s$.
Thus, combining measurements of the oscillation frequency of
both $\bdmix$ and $\bsmix$ mixing translates into a measurement of
the ratio $|V_{td}| / |V_{ts}|$.

  Experimentally, a measurement of the time dependence of $\bmix$
mixing requires three ingredients: (i) the $B$ decay proper time has
to be reconstructed, (ii) the $B$ flavor at production
(initial state $t = 0$) needs to be determined, as well as (iii) the $B$
flavor at decay (final state $t = t_{\rm{decay}}$).
At SLD, the time dependence of $\bdmix$ mixing has been measured using
four different methods. All four use the same initial state tagging
but differ by the method used to either reconstruct the $B$ decay or
tag its final state.

  Initial state tagging takes advantage of the large polarization-dependent
forward backward asymmetry in $\Zbb$ decays
\begin{equation}
\tilde{A}_{FB}(\cos\theta_T) = 2 A_b~{{A_e - P_e}\over{1 - A_e P_e}}
        ~{{\cos\theta_T}\over{1+\cos^2\theta_T}},
\end{equation}
where $A_b = 0.94$ and $A_e = 0.15$.
This only requires knowledge
of the electron beam polarization $P_e$ and the cosine of the angle between
the thrust axis direction $\hat{T}$ and the electron beam direction,
$\cos\theta_T$.
For left- (right-) handed electrons and
forward (backward) $B$ decay vertices, the initial quark is tagged as
a $b$ quark; otherwise, it is tagged as a $\overline{b}$ quark.
The initial state tag can be augmented by using a momentum-weighted
track charge in the hemisphere opposite
that of the reconstructed $B$ vertex, defined as
\begin{equation}
  Q_{jet} =
  \sum Q_i \left|\vec{p_i} \cdot \hat{T}\right|^\kappa
  \mbox{sign}\left(\vec{p_i} \cdot \hat{T} \right),
\end{equation}
where $\vec{p_i}$ is the three-momentum of track $i$ and $Q_i$ its
charge, and $\kappa = 0.5$.
\begin{figure}[b]
  \vspace{0 mm}
  \centering
  \epsfxsize7.4cm
  \leavevmode
  \epsfbox{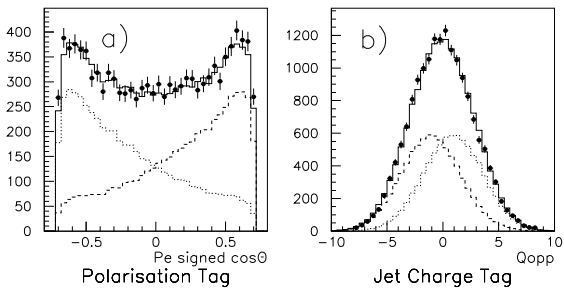}
  \vspace{-7mm}
  \caption{Distributions of (a) polarization-signed $\cos\theta_T$
           and (b) opposite hemisphere jet charge for data (points)
           and MC (solid line). The MC $b$ and $\bar{b}$ components
           are shown with dotted and dashed lines respectively.}
  \label{Fig_inittag}
\end{figure}
Figure~\ref{Fig_inittag} shows the distributions of $\cos\theta_T$ signed
by the electron beam helicity and opposite hemisphere jet charge.
Clear separation between initial $b$ and $\bar{b}$ quarks is observed.
These two tags are
combined to yield an initial state tag with 100\% efficiency and
effective average right-tag probability of 84\%
(for $\langle P_e \rangle = 77\%$).

  The first two $\bdmix$ mixing analyses~\cite{bmixkadi}
use topological vertexing
to select the tracks from the $B$ decay and measure
its decay length. A sample of 16803 vertices is selected after
requiring the mass of all tracks in the vertex, corrected for the amount
of missing transverse momentum, $M > 2$ GeV/c$^2$
(no explicit cut is placed on the decay length).
The first analysis uses charged kaons from the $B$ decay chain to
tag the final state. This tag relies on the fact that most $B$ decays
occur via the dominant $b \to c \to s$ transition.
Thus, $K^-$ ($K^+$) tags $\overline{\bd}$ ($\bd$) decays.
The fraction of charged kaons produced with the right sign
has been measured to be $(82 \pm 5)\%$ in $\bd$ decays~\cite{ArgusK}.
Charged kaons are identified with the Cherenkov Ring Imaging Detector,
using both liquid and gas radiators to cover most of the kaon momentum
range: 0.8 to 25 GeV/c.
The rate of pion misidentification as a function of momentum is
calibrated from the data using a pure sample of pions from $K^0_s$ decays.
The kaon tag yields a sample of 5694 decays with a correct tag probability
of 77\% for $\bd$ decays.

\begin{figure}[htb]
  \vspace{-5mm}
  \centering
  \epsfxsize8.3cm
  \leavevmode
  \epsfbox{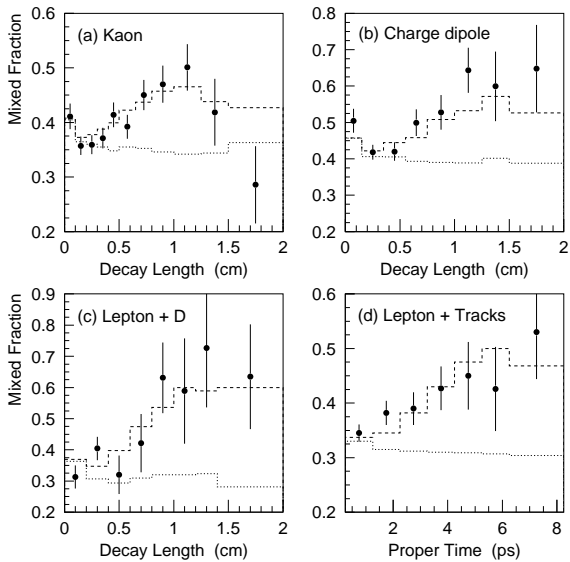}
  \vspace{-8mm}
  \caption{Distributions of the fraction of decays tagged as mixed as
           a function of decay length or proper time for data (points)
           and best fit MC (dashed histograms) for the various
           analyses: (a) charged kaon, (b) charge dipole,
           (c) lepton + $D$, and (d) lepton + tracks.
           The dotted histograms correspond to MC distributions with
           no $\bdmix$ mixing.}
  \label{Fig_bdmix}
\end{figure}

  The time dependence of $\bdmix$ mixing is measured
from the fraction of decays tagged as mixed as a function of
decay length.
A decay is tagged as mixed if the initial and final
state tags disagree.
A binned $\chi^2$ fit is performed by comparing the distributions of
the mixed fraction as a function of decay length obtained for the
data and the MC for a range of $\Delta m_d$ values.
Figure~\ref{Fig_bdmix}(a) shows the mixed fraction distribution for
the charged kaon analysis.
The fit yields a frequency of
$\Delta m_d = 0.580 \pm 0.066(\mbox{stat}) \pm 0.075(\mbox{syst})$ ps$^{-1}$
with a $\chi^2$/dof = 10.2/10.
The main contributions to the systematic error arise from uncertainties
in the $\pi \to K$ misidentification calibration from the data,
in the rate of right-sign kaon production in $\bu$ and $\bd$ decays,
and in the dependence of the fit results on binning and fit range,
as summarized in Table~\ref{Tbl_bdmixsyst}.

\begin{table*}[hbt]
\setlength{\tabcolsep}{1.5pc}
\newlength{\digitwidth} \settowidth{\digitwidth}{\rm 0}
\catcode`?=\active \def?{\kern\digitwidth}
\caption{Systematic uncertainties for the different $\Delta m_d$ measurements
 (in ps$^{-1}$).}
\label{Tbl_bdmixsyst}
\begin{tabular*}{\textwidth}{@{}l@{\extracolsep{\fill}}rrrr}
\hline
 Analysis & Kaon & Charge Dipole & Lepton+$D$ & Lepton+Tracks \\
  \cline{1-1} \cline{2-5}
 Detector simulation & 0.036 & 0.010 & 0.020 & 0.013 \\
 Physics modeling    & 0.048 & 0.027 & 0.024 & 0.032 \\
 Fit systematics     & 0.045 & 0.026 & 0.038 &  --   \\
  \cline{1-1} \cline{2-5}
 Total               & 0.075 & 0.039 & 0.049 & 0.035 \\
 \hline
\end{tabular*}
\end{table*}

  The second analysis exploits the $B \to D$ cascade charge structure to
tag the final state. To enhance the $\bd$ fraction,
we require the vertex charge $Q = 0$.
This requirement also improves the probability of correctly assigning
all of the $\bd$ decay tracks.
The direction of the vertex axis is adjusted to minimize the impact
parameter sum of the tracks in the vertex and the mean track impact parameter
is required to be less than 50 $\mu$m at this minimum.
A sample of 3291 decays satisfies the selection criteria.
The ``charge dipole'' $\delta q$ of the vertex is then defined
as the relative displacement between the weighted mean location\str
\vfill\pagebreak\noindent
$L_i$ of the positive tracks and of the negative tracks:
 $\delta q = (\sum^+ w_i L_i) / (\sum^+ w_i)-(\sum^- w_i L_i) / (\sum^- w_i)$,
where the first (second) term is a sum over all positive (negative) tracks
in the vertex. The weight $w_i$ for each track $i$ is
inversely proportional to the
uncertainty on the quantity $L_i$.
The correct tag probability increases with the magnitude of $\delta q$
and reaches a maximum of 84\% at large $|\delta q|$.
A fit to the mixed fraction distribution as a function
of decay length [Fig.~\ref{Fig_bdmix}(b)] yields
$\Delta m_d = 0.561 \pm 0.078(\mbox{stat}) \pm 0.039(\mbox{syst})$ ps$^{-1}$
with a $\chi^2$/dof = 8.8/7.
The main contributions to the systematic error come from MC statistics
and fit systematics (Table~\ref{Tbl_bdmixsyst}).

  The next two analyses select semileptonic decays. The first of these
(lepton~+~$D$, Ref.~\cite{bmixldvx}) is identical to that used to measure
the $\bu$ and $\bd$ lifetimes.
As for the charge dipole analysis, a set
of neutral vertices is selected.
The charge of the lepton tags the $\bd$/$\overline{\bd}$ flavor
at decay with a correct tag probability of 85\%.
As for the above two analyses, a fit to the mixed fraction
distribution [Fig.~\ref{Fig_bdmix}(c)] yields
$\Delta m_d = 0.452 \pm 0.074(\mbox{stat}) \pm 0.049(\mbox{syst})$ ps$^{-1}$
with a $\chi^2$/dof = 7.8/7.
The systematic error is dominated by MC statistics and fit
systematics (Table~\ref{Tbl_bdmixsyst}).

  The last analysis (lepton$+$tracks, Ref.~\cite{bmixltrk})
selects semileptonic decays by
identifying electrons and muons with high transverse
momentum $p_T > 0.8$ GeV/c with respect to the nearest jet axis.
This enhances the fraction of $\Zbb$ events and allows
for the use of a fully inclusive vertexing technique.
The $B$ decay vertex is estimated by computing an average intersection
point between the lepton trajectory and all well-measured tracks in the jet,
each track being weighted according to its probability to originate
from the decay of a short-lived heavy hadron.
Here, the $B$ decay
proper time is reconstructed by estimating the $B$ hadron momentum
based on track and energy clusters in the calorimeter.
The final sample contains 2609 semileptonic decay candidates with
an estimated $B$ hadron purity of 93\% and correct final state tag probability
of 85\%.

For this analysis, the value of $\Delta m_d$ is extracted from
an unbinned maximum likelihood analysis with parameterizations
estimated from the MC. The fit yields
$\Delta m_d = 0.520 \pm 0.072(\mbox{stat}) \pm 0.035(\mbox{syst})$ ps$^{-1}$
and the corresponding mixed fraction distribution is shown
in Fig.~\ref{Fig_bdmix}(d).
Main systematic uncertainties are presented in Table~\ref{Tbl_bdmixsyst}.

  The preliminary results from the four analyses have been
combined, taking into account statistical and systematic correlations,
to produce the following SLD average:
\begin{equation}
  \Delta m_d = 0.525 \pm 0.043(\mbox{stat}) \pm 0.037(\mbox{syst})
   \mbox{~ps}^{-1}.
\end{equation}
This average is consistent with the world average value of
$0.466 \pm 0.018$~\cite{Gibbons}.

\section{SUMMARY AND FUTURE}
\label{Sec_sum}

  Using a sample of $\sim 150,000$ hadronic $Z^0$ decays collected
between 1993 and 1995, the SLD Collaboration has produced precise
$\bu$ and $\bd$ lifetime measurements, as well as
measurements of the time dependence of $\bdmix$ mixing.
  In 1996, SLD installed an improved CCD Vertex Detector.
This new detector allows for significant improvements in resolution.
In particular, the decay length resolution improves by roughly
a factor of two. SLD is looking forward to performing many exciting
$B$ physics measurements over the next few years.

%
%
%
\footnotesize
  \def\iADEL{$^{(1)}$}
  \def\iBOL{$^{(2)}$}
  \def\iBU{$^{(3)}$}
  \def\iBRUN{$^{(4)}$}
  \def\iUCSB{$^{(5)}$}
  \def\iUCSC{$^{(6)}$}
  \def\iCIN{$^{(7)}$}
  \def\iCSU{$^{(8)}$}
  \def\iCOLO{$^{(9)}$}
  \def\iCOL{$^{(10)}$}
  \def\iFER{$^{(11)}$}
  \def\iFRA{$^{(12)}$}
  \def\iILL{$^{(13)}$}
  \def\iLBL{$^{(14)}$}
  \def\iMIT{$^{(15)}$}
  \def\iMASS{$^{(16)}$}
  \def\iMISS{$^{(17)}$}
  \def\iMOSC{$^{(18)}$}
  \def\iNAG{$^{(19)}$}
  \def\iOREG{$^{(20)}$}
  \def\iPAD{$^{(21)}$}
  \def\iPERU{$^{(22)}$}
  \def\iPISA{$^{(23)}$}
  \def\iRUT{$^{(24)}$}
  \def\iRAL{$^{(25)}$}
  \def\iSOGANG{$^{(26)}$}
  \def\iSOONG{$^{(27)}$}
  \def\iSLAC{$^{(28)}$}
  \def\iTENN{$^{(29)}$}
  \def\iTOH{$^{(30)}$}
  \def\iVAND{$^{(31)}$}
  \def\iWASH{$^{(32)}$}
  \def\iWISC{$^{(33)}$}
  \def\iYALE{$^{(34)}$}
  \def\dead{$^{\dag}$}
  \def\andgen{$^{(a)}$}
  \def\andper{$^{(b)}$}
%
%
\begin{center}
\mbox{$^{\ast}$K. Abe                 \unskip,\iNAG}
\mbox{K. Abe                 \unskip,\iTOH}
\mbox{T. Akagi               \unskip,\iSLAC}
\mbox{N.J. Allen             \unskip,\iBRUN}
\mbox{W.W. Ash               \unskip,\iSLAC$^\dagger$}
\mbox{D. Aston               \unskip,\iSLAC}
\mbox{K.G. Baird             \unskip,\iMASS}
\mbox{C. Baltay              \unskip,\iYALE}
\mbox{H.R. Band              \unskip,\iWISC}
\mbox{M.B. Barakat           \unskip,\iYALE
% ...
}
\it
  \vskip \baselineskip                   
  \centerline{(The SLD Collaboration)}   
  \vskip \baselineskip                   
%
%
%
  \iADEL
     Adelphi University,
     Garden City, New York 11530 \break
  \iBOL
     INFN Sezione di Bologna,
     I-40126 Bologna, Italy \break
  \iBU
     Boston University,
     Boston, Massachusetts 02215 \break
  \iBRUN
     Brunel University,
     Uxbridge, Middlesex UB8 3PH, United Kingdom \break
  \iUCSB
     University of California at Santa Barbara,
     Santa Barbara, California 93106 \break
  \iUCSC
     University of California at Santa Cruz,
     Santa Cruz, California 95064 \break
  \iCIN
     University of Cincinnati,
     Cincinnati, Ohio 45221 \break
  \iCSU
     Colorado State University,
     Fort Collins, Colorado 80523 \break
  \iCOLO
     University of Colorado,
     Boulder, Colorado 80309 \break
  \iCOL
     Columbia University,
     New York, New York 10027 \break
  \iFER
     INFN Sezione di Ferrara and Universit\`a di Ferrara,
     I-44100 Ferrara, Italy \break
  \iFRA
     INFN  Lab. Nazionali di Frascati,
     I-00044 Frascati, Italy \break
  \iILL
     University of Illinois,
     Urbana, Illinois 61801 \break
  \iLBL
     Lawrence Berkeley Laboratory, University of California,
     Berkeley, California 94720 \break
  \iMIT
     Massachusetts Institute of Technology,
     Cambridge, Massachusetts 02139 \break
  \iMASS
     University of Massachusetts,
     Amherst, Massachusetts 01003 \break
  \iMISS
     University of Mississippi,
     University, Mississippi  38677 \break
  \iMOSC
    Moscow State University,
    Institute of Nuclear Physics
    119899 Moscow, Russia    \break
  \iNAG
     Nagoya University,
     Chikusa-ku, Nagoya 464 Japan  \break
  \iOREG
     University of Oregon,
     Eugene, Oregon 97403 \break
  \iPAD
     INFN Sezione di Padova and Universit\`a di Padova,
     I-35100 Padova, Italy \break
  \iPERU
     INFN Sezione di Perugia and Universit\`a di Perugia,
     I-06100 Perugia, Italy \break
  \iPISA
     INFN Sezione di Pisa and Universit\`a di Pisa,
     I-56100 Pisa, Italy \break
  \iRUT
     Rutgers University,
     Piscataway, New Jersey 08855 \break
  \iRAL
     Rutherford Appleton Laboratory,
     Chilton, Didcot, Oxon OX11 0QX United Kingdom \break
  \iSOGANG
     Sogang University,
     Seoul, Korea \break
  \iSOONG
     Soongsil University,
     Seoul, Korea  156-743 \break
  \iSLAC
     Stanford Linear Accelerator Center, Stanford University,
     Stanford, California 94309 \break
  \iTENN
     University of Tennessee,
     Knoxville, Tennessee 37996 \break
  \iTOH
     Tohoku University,
     Sendai 980 Japan \break
  \iVAND
     Vanderbilt University,
     Nashville, Tennessee 37235 \break
  \iWASH
     University of Washington,
     Seattle, Washington 98195 \break
  \iWISC
     University of Wisconsin,
     Madison, Wisconsin 53706 \break
  \iYALE
     Yale University,
     New Haven, Connecticut 06511 \break
  \dead
     Deceased \break
  \andgen
     Also at the Universit\`a di Genova \break
  \andper
     Also at the Universit\`a di Perugia \break
\rm
%

\end{center}



\begin{thebibliography}{9}
\bibitem{Rb} K.~Abe {\it et al.},
             Phys. Rev. D {\bf 53}, 1023 (1996).
\bibitem{Bigi} I. I.~Bigi~{\it et~al.}, in {\sl B Decays},
 ed. S.~Stone (World Scientific, New York, 1994), p.~132.
\bibitem{blife} K.~Abe {\it et al.},
 to appear in Phys. Rev. Lett. {\bf 79} (1997).
\bibitem{ZVTOP} D.~Jackson, Nucl.~Instrum. Methods A {\bf 388}, 247 (1997).
\bibitem{Gibbons}
 L. K.~Gibbons, {\it Status of Weak Quark Mixing},
 UR-1494, April 1997.
\bibitem{bmixkadi} K.~Abe {\it et al.},
 SLAC-PUB-7230, July 1996.
\bibitem{ArgusK} H.~Albrecht {\it et al.},
                 Z. Phys. C {\bf 62}, 371 (1994).
\bibitem{bmixldvx} K.~Abe {\it et al.},
 SLAC-PUB-7229, July 1996.
\bibitem{bmixltrk} K.~Abe {\it et al.},
 SLAC-PUB-7228, July 1996.

\end{thebibliography}
\end{document}